\documentclass{article}
\usepackage[T1]{fontenc} 
\usepackage[utf8]{inputenc} 
\usepackage{ismir,amsmath,cite,url}
\usepackage{underscore}
\usepackage{graphicx}
\usepackage{color}
\usepackage{booktabs}  
\usepackage{enumitem}

\usepackage[bookmarks=false,hidelinks]{hyperref}

\def \mdtkurl {https://www.github.com/JamesOwers/midi_degradation_toolkit}
\def \acmeurl {https://www.github.com/JamesOwers/acme}

\title{The MIDI Degradation Toolkit:\\Symbolic Music Augmentation and Correction}

\threeauthors
  {Andrew McLeod} {EPFL \\ {\tt\small andrew.mcleod@epfl.ch}}
  {James Owers} {University of Edinburgh \\ {\tt\small james.owers@ed.ac.uk}}
  {Kazuyoshi Yoshii} {Kyoto University \\ {\tt\small yoshii@kuis.kyoto-u.ac.jp}}

\begin{document}

\maketitle
\begin{abstract}
In this paper, we introduce the MIDI Degradation Toolkit (MDTK), containing functions which take as input a musical excerpt (a set of notes with pitch, onset time, and duration), and return a ``degraded'' version of that excerpt with some error (or errors) introduced. Using the toolkit, we create the Altered and Corrupted MIDI Excerpts dataset version 1.0 (ACME v1.0), and propose four tasks of increasing difficulty to detect, classify, locate, and correct the degradations. We hypothesize that models trained for these tasks can be useful in (for example) improving automatic music transcription performance if applied as a post-processing step. To that end, MDTK includes a script that measures the distribution of different types of errors in a transcription, and creates a degraded dataset with similar properties. MDTK's degradations can also be applied dynamically to a dataset during training (with or without the above script), generating novel degraded excerpts each epoch. MDTK could also be used to test the robustness of any system designed to take MIDI (or similar) data as input (e.g. systems designed for voice separation, metrical alignment, or chord detection) to such transcription errors or otherwise noisy data. The toolkit and dataset are both publicly available online, and we encourage contribution and feedback from the community.
\end{abstract}
\section{Introduction}
\label{sec:intro}
Music language models (MLMs) have been the subject of much research in recent years. In the most general terms, their goal is to learn the structure of a typical piece of music, usually in symbolic form, as either a piano roll or a (monophonic or polyphonic) sequence of notes. Such models can be designed either as a stand-alone system (i.e. to perform a specific task such as voice separation, metrical alignment, or chord detection), or as part of an automatic music transcription (AMT) system along with an acoustic model.

In AMT systems, MLMs have thus far led to only small increases in performance compared to state-of-the-art acoustic models by themselves \cite{Kelz2016}. One possible reason is that such MLMs are typically run at the frame-level\footnote{It is debatable as to whether frame-based models should be called ``language'' models, since they do not work at a step related to the language (e.g. musical notes or beats), but rather the frames of the acoustic model. However, such a distinction is not the focus of this work.}, rather than at the note-level or the beat-level \cite{Ycart2017}. Regardless, even beat- or note-level MLMs have not led to very large improvements by themselves (e.g. \cite{app8030470, ycart2018polyphonic}). One approach to solving this issue has been proposed in \cite{McLeod:19b}, where a separate ``blending model'' is used to combine the acoustic model with the MLM. The blending model leads to a small but significant increase in performance over using the acoustic model only.

Another possible reason for their minimal improvement is that such MLMs are not directly trained to solve the task at hand---to correct errors produced by the acoustic model. That is, they are not \textit{discriminative} models taking data with errors as input and producing the correct transcription as output. Instead, they are typically trained to model the distribution of clean (usually MIDI) data, and used to alter the probabilistic predictions of the acoustic model. The integration of such an MLM into an AMT system usually involves searching through a large space of possible output transcriptions. One potential solution to this problem (at least when using an RNN-based MLM), is to train the model with scheduled sampling \cite{Bengio2015}, which uses its own (noisy) outputs during training, teaching it to recover from such mistakes. In fact, the MLM from \cite{McLeod:19b} is trained using scheduled sampling. However, this training strategy is only designed to allow the MLM to recover after a mistake, rather than to recognize and correct a mistake directly.

Training a discriminative model which ``cleans'' the output of an acoustic model is only feasible in the presence of a dataset mapping degraded data to clean data. Whilst this dataset could be produced by running an acoustic model on a dataset mapping audio to the correct transcription, such datasets are small relative to the amount of clean MIDI data available elsewhere. Our MDTK package allows the user to take any clean data and degrade it to have musical errors of their choosing. The pool of clean MIDI data is many orders of magnitude larger than that which maps audio to transcription data. For example, MAESTRO \cite{hawthorne2018enabling} has aligned MIDI and audio data of \textasciitilde{}1\,300 performances totalling \textasciitilde{}200 hours. In comparison, the Lakh MIDI Dataset \cite{raffel2016learning} comprises \textasciitilde{}175\,000 MIDI files\footnote{The true number of files is slightly smaller than this as it is known that some of these MIDI files are corrupted.} totalling \textasciitilde{}9\,000 hours. This is over 40 times the size, and additionally spans diverse genres. Using a dataset such Lakh MIDI, MDTK allows for the creation of datasets large enough to make the direct discriminative task feasible. In addition, as we will discuss later in Section \ref{subsec:tools}, there is no need to restrict learning capability by explicitly creating a degraded dataset: MDTK's \texttt{\mbox{Degrader}} objects can be used to degrade clean input dynamically when loading it into the model, thus providing on-the-fly data augmentation, enabling the model to be trained on a degraded dataset which is essentially unlimited in size.

This process is analogous to performing learned data augmentation---MDTK makes the discriminative task of correcting errors feasible by increasing the effective size of the dataset. Data augmentation has proved essential in other fields. In \cite{Cubuk2019CVPR}, the authors advocate the automated application of data augmentation for the ImageNet task \cite{imagenetcvpr09}, a classification task for image data. They find that by automatically tuning the type of data augmentation they apply for each task, they can attain a significant improvement over the state-of-the-art. In \cite{antoniou2017data}, the authors explicitly investigate the effects of generating augmented data in low-data regimes, advocating the use of learned generators---essentially what MDTK's \texttt{\mbox{Degrader}} objects are---using GANs. Finally, in \cite{7829341}, the authors solve their low-data regime issue for environmental sound classification by using data augmentation, finding that performing augmentations such as pitch shifting and time stretching leads to a 6 percentage point boost in classification accuracy. MDTK enables similar such data augmentation techniques to be performed easily for AMT.

For non-AMT tasks, standalone MLMs typically take as input MIDI files and output some alignment or label, depending on the task. To our knowledge, the robustness of these MLMs to noisy or incorrect data is rarely if ever analysed. This is not necessarily an important factor when clean MIDI files are used as input, but when such a MIDI file is the result of noisy process such as AMT or Optical Music Recognition (OMR; e.g. \cite{van2017optical}), a model's robustness to noise becomes an important piece of information.

We propose that both of these shortcomings---poor AMT post-processing, and that MLMs' robustness to noise has not been analysed---can be addressed using excerpts of music to which noise is added. In an AMT system, a post-processing model which is trained directly to identify and correct similar noise should be better able to correct noisy acoustic model outputs than a generic MLM. Likewise, the robustness of a standalone MLM to noisy input can be analyzed with such noisy data, allowing the MLM to be evaluated for its potential usefulness in downstream tasks such as those involved in creating a complete piece of sheet music given an audio signal.

In this paper, we introduce the MIDI Degradation Toolkit (MDTK), a set of tools to easily introduce controlled noise into excerpts automatically extracted from a set of MIDI files. MDTK is similar to the Audio Degradation Toolbox \cite{matthias2013a} for audio, but to the authors' knowledge, ours is the first toolkit of its kind for MIDI data. The controlled noise includes (1) shifting the pitch of a note; (2) lengthening, shortening, or shifting a note in time; (3) adding or removing a note; and (4) splitting or joining notes.

We also introduce the Altered and Corrupted MIDI Excerpts dataset version 1.0 (ACME v1.0), containing MIDI excerpts which have been degraded (and some which have not been degraded) using the toolkit, and four new tasks of increasing difficulty: to (1) detect whether each excerpt has been degraded; (2) if so, classify what degradation has been applied and (3) locate where a degradation has taken place; and (4) recover the original excerpt.

We present a simple baseline model for each task and analyse its performance. These baselines are provided as an easy starting point for researchers wanting to attempt our proposed tasks or post-process their own AMT data. We provide evaluation metrics for assessment and postulate that, if high performance were achieved, we would be able to improve AMT output using models trained for these tasks. We can easily swap out ACME v1.0 for a dataset matching the errors for a specific AMT system using a provided script.

\section{The Toolkit}
\label{sec:mdtk}
The MIDI Degradation Toolkit (MDTK) is a python package, installable with pip, which can be used to introduce errors to MIDI excerpts. The code is released open source under an MIT License, and is available online. We encourage feedback and contribution from the community in its continued development.

Internally, MDTK stores each excerpt as a set of notes in a Pandas \cite{pandas} DataFrame with columns pitch (MIDI pitch, with C4=60), onset (the onset time of the note, in milliseconds), track, and dur (the duration of the note, in milliseconds), all integers. It contains functionality to load an excerpt from a MIDI file (using pretty\_midi \cite{prettymidi}), as well as to read from and write to a CSV file.

\subsection{Degradations}
\label{subsec:deg}
Each degradation provided in MDTK takes as input a pandas DataFrame of an excerpt of music, and returns a DataFrame with the given degradation. Some degradations (e.g. removing a note from an empty excerpt) are not always possible. In such cases, a warning is printed and None is returned. Care is also taken to ensure that no overlaps on the same pitch are introduced by a degradation. There are a total of 8 degradations in MDTK, each of which is described below.

The \texttt{\mbox{pitch_shift}} degradation changes the pitch of a random note. By default, the new pitch is chosen uniformly at random from all possible pitches (a minimum and maximum pitch can be given, and the valid range defaults to 21--108 inclusive). It can also be drawn from a weighted distribution of intervals around the original pitch, for example to emphasize octave errors from overtones. We also include a flag to force the new pitch to align with the pitch of some other note in the excerpt, to reduce out-of-key shifts, if desired.

Three degradations shift a random note in time in some way: \texttt{\mbox{onset_shift}} changes the note's onset time, leaving its offset time unchanged; \texttt{\mbox{offset_shift}} changes the note's offset time, leaving its onset time unchanged; and \texttt{\mbox{time_shift}} changes the note's onset and offset times by the same amount, leaving its duration unchanged. For all of these degradations, care is taken to ensure that the shifted note does not lie outside the excerpt's initial time range. A minimum and maximum resulting duration can be specified, as well as a minimum and maximum shift amount. We also include flags to align some combination of the shifted note's onset or duration with those of other notes from the excerpt, ensuring the note lies on some metrical grid, if desired.

Two degradations can be used to either add a random note to an excerpt (\texttt{\mbox{add_note}}), or remove a random note from an excerpt (\texttt{\mbox{remove_note}}). Flags to align an added note's pitch, onset, or duration to those of existing notes are included.

Two degradations can be used either to split a note into multiple shorter consecutive notes or to combine consecutive notes at the same pitch into a single longer note. Specifically, \texttt{\mbox{split_note}} will cut a random note into some number of consecutive notes of shorter duration (the first of which begins at the original note's onset time and the last of which ends at the original note's offset time). By default the note is split into two shorter notes, but this---as well as a minimum allowable duration for the resulting notes---can be set with a parameter. Similarly, \texttt{\mbox{join_notes}} takes two or more consecutive notes at the same pitch (with a maximum allowable gap---set with a parameter---allowed between them), and joins them into a single note with onset time equal to that of the first note and offset time equal to that of the last.

\subsection{Other Tools}
\label{subsec:tools}
\subsubsection{Dynamically degrading clean data}
MDTK includes the \texttt{\mbox{Degrader}} class, which can be used to degrade excerpts dynamically. When instantiating a \texttt{\mbox{Degrader}} object, the proportion of excerpts that should remain undegraded is set with a parameter (which can be 0). The probability of each degradation being performed on an excerpt (if it is to be degraded) can also be set at this time. Then, each time \texttt{Degrader.degrade(excerpt)} is called, a randomly degraded version of the input excerpt is generated according to the proportions set during object creation. The \texttt{\mbox{Degrader}} class can be easily inserted into any model training procedure in order to dynamically create new degraded excerpts during each epoch, dramatically increasing the amount of data available for training.

\subsubsection{Automatically matching model errors}
MDTK includes a \texttt{\mbox{measure_errors.py}} script, which can be used to estimate the types of errors (specifically, as degradations) typical to a particular AMT system, given a set of transcriptions and ground truths from that system. Note that there is no unique set of degradations which reproduce the errors that a transcription system has made (e.g., any \texttt{shift} degradation can be trivially replaced by a \texttt{\mbox{remove_note}} and an \texttt{\mbox{add_note}}). We make no claim that the degradations found by the script correspond to the exact causes of the errors made by the AMT system. Rather, only that the distribution of degradations produces excerpts with similar properties to those transcribed by that system. Nonetheless, the script finds what we believe are a \textit{reasonable} set of degradations to have produced those errors using a simple heuristic-based approach. Notes are first matched as correct if possible (same pitch, and onset and offset within a changeable threshold), and the remaining notes are checked for the various degradations in the following order: (1) \texttt{\mbox{join_notes}} and \texttt{\mbox{split_note}}, either of which may include an additional \texttt{\mbox{offset_shift}} or \texttt{\mbox{onset_shift}}; (2) \texttt{\mbox{offset_shift}}, if the pitch and onset time match; (3) \texttt{\mbox{onset_shift}}, if the pitch and offset time match; (4) \texttt{\mbox{time_shift}}, but only if the transcribed note overlaps the position of the corresponding ground truth note; and (5) \texttt{\mbox{pitch_shift}}, which must match in onset time, although an additional \texttt{\mbox{offset_shift}} can be added. Finally, any remaining unmatched notes are counted as \texttt{\mbox{add_note}} and \texttt{\mbox{remove_note}}.

The output of the script is a json file containing the estimated proportion of each degradation in the given set of transcriptions. It does not yet include values for the various degradation parameters (though this is planned for a future update to MDTK). This output file can be used, for example, to create a custom-tuned, static, degraded dataset for training a model. However, the two tools can also be combined in powerful ways. By passing this json file to the \texttt{\mbox{Degrader}} constructor, a \texttt{\mbox{Degrader}} object can be instantiated that generates degradations exactly matching the estimated proportions. This could then be used to train a model to correct the errors of that specific AMT system using a relatively small amount of raw data.

\section{The Dataset}
\label{sec:acme}
\subsection{ACME version 1.0}

The Altered and Corrupted MIDI Excerpts dataset v1.0 (ACME v1.0) is a dataset of 5 second excerpts with degradations implemented by MDTK. It is not intended to emulate the errors of any specific AMT system, but rather serve as a starting point for the modelling tasks we introduce below.

The dataset is taken from two sources: (1) the piano-midi dataset\footnote{\url{http://www.piano-midi.de}}, which contains 328 MIDI files of pseudo-live performance\footnote{The files are quantized and beat-aligned, but their tempo curves were manually edited by their creator to sound more like live performance.} piano pieces of various styles (generally classical); (2) the 22194 primes from the small, medium, and large sections of the monophonic and polyphonic Patterns for Prediction Development Datasets (PPDD-Sep2018)\footnote{\url{https://www.music-ir.org/mirex/wiki/2019:Patterns_for_Prediction}}, which contain excerpts drawn randomly from the Lakh MIDI Dataset (LMD) \cite{raffel2016learning}.

We remove track information, flattening each excerpt to a single track, simplifying the modelling tasks\footnote{The use of tracks is not standard our different data sources.}; analysis of multi-track MIDI files will be addressed in future work. We then fix any pair of overlapping notes of the same pitch by cutting the first note at the onset time of the second. We additionally set the offset time of the second note to the maximum of the original offset times of the two notes, such that no sustain is removed.

Once this pre-processing is complete, we select a \textasciitilde 5 second excerpt from each piece by choosing a random note and all notes beginning in the subsequent 5 seconds, but require that at least 10 notes be present. The excerpt ends when the last held note is released. This duration is approximately 2 bars for most songs so is small enough for the models proposed in section \ref{sec:tasks} to train quickly. We degrade $\frac{8}{9}$ of the excerpts, selecting the degradation uniformly at random from the set of 8 defined degradations, and leave the remaining $\frac{1}{9}$ undegraded. For ACME v1.0, we use default parameter settings for all degradations, although we intend to investigate the effect of different settings in future work (and future releases of ACME datasets).

The excerpts and degraded excerpts are split randomly into training, validation, and test sets of proportion $0.8$, $0.1$, and $0.1$, creating the official splits for ACME v1.0. The canonical form is available online as a set of CSV files. Additionally, the MDTK package includes the \mbox{\texttt{make\_dataset.py}} script which we used to create the dataset from scratch---including the automatic downloading of the raw data---and thus serves as a record of how it was created.

\subsection{Custom Dataset Creation}
The \mbox{\texttt{make\_dataset.py}} script can also be used to generate an ACME-style dataset from a user-provided set of MIDI or CSV files. The user can specify custom sizes for the excerpts, a custom distribution of the various degradations, as well as custom parameters for each. The script can be given the json output of the \texttt{\mbox{measure_errors.py}} script in order to match the properties of the generated dataset with those measured from an AMT result. Alternatively, a user can simply choose to degrade individual excerpts from their own training set by calling MDTK during the training process, either manually or randomly using the \texttt{\mbox{Degrader}} class.

\section{Proposed Tasks}
\label{sec:tasks}

\subsection{Motivation}
These tasks are performed on ACME v1.0, and proposed in lieu of taking existing AMT systems and measuring their improvement when trained with the assistance of MDTK. It is proposed that the \textit{output} of arbitrary AMT systems could be improved with models that can solve these tasks.

For instance, we could use a model trained to classify the error contained within a given excerpt to call out for human intervention. We could also train models to perform the actual fix; however, we show that, with the models we have chosen for our baseline, this problem is far from solved.

\subsection{Tasks}
\begin{figure}
    \centering
    \includegraphics[width=0.48\textwidth,trim={0.3cm 3cm 0.3cm 3.2cm},clip]{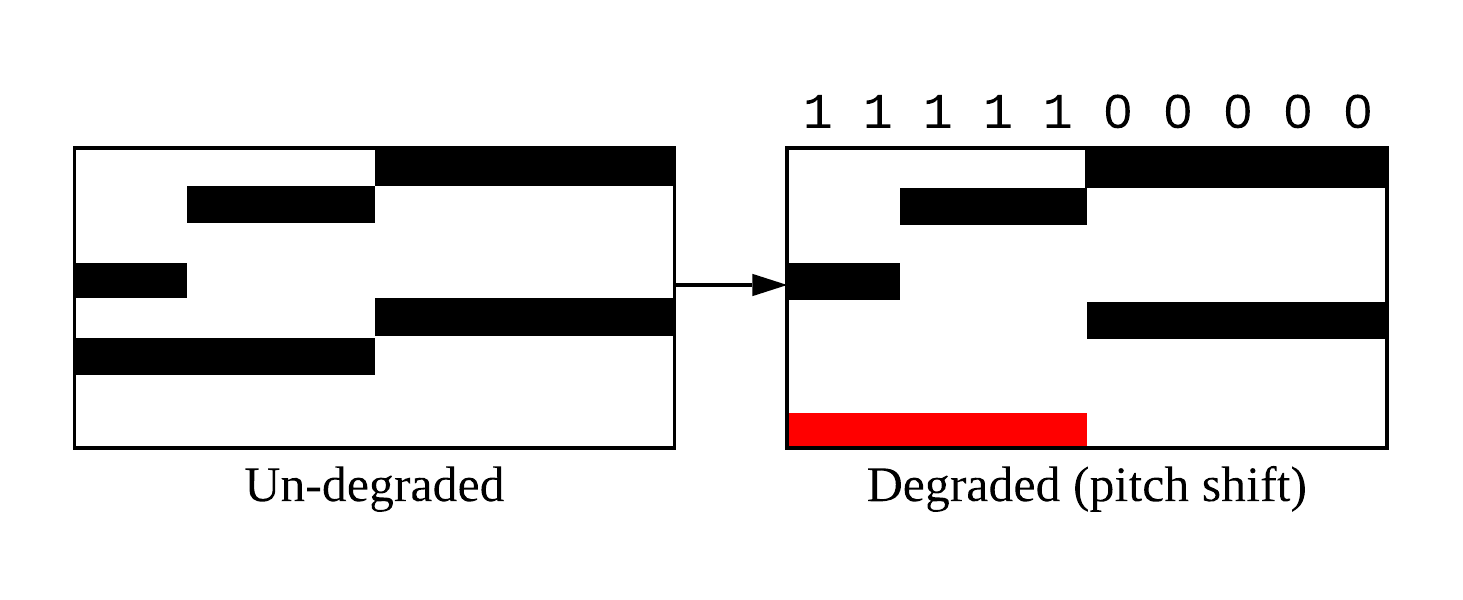}
    \caption{Example piano rolls of a clean excerpt (left) being degraded with pitch\_shift (right), including the labels for Error Location (top right).}
    \label{fig:example}
\end{figure}

We propose four tasks of increasing difficulty. Figure \ref{fig:example} shows a simple toy data point which has been pitch shifted (changed note in red). We will use it as an example when necessary throughout this section. We should note that the tasks we introduce here are not in any way trivial, but represent significant steps towards successful AMT post-processing.

\textbf{1. Error Detection}: detect whether a given excerpt has been degraded. This is a binary classification task with a skewed distribution: $\frac{8}{9}$ excerpts are degraded (the positive class), and $\frac{1}{9}$ are not degraded (the negative class). We evaluate performance using F-measure but, since the negative class is the minority, for the purposes of F-measure evaluation, we treat those as positives. Thus, a model which always outputs ``degraded'' achieves a ``reverse F-measure'' of $0.00$ (with precision and recall both 0) rather than its F-measure of $0.94$ (with precision $\frac{8}{9}$ and recall $1$).

\textbf{2. Error Classification}: specify which degradation (if any) was performed on each excerpt. This is a multi-class classification problem, and since ACME v1.0 contains a uniform distribution of each class, we evaluate performance using accuracy and a confusion matrix to show specific error tendencies for each degradation.

\textbf{3. Error Location}: assign a binary label to each (40 ms) frame of input identifying whether it contains an error i.e. whether this frame contains a degradation. We evaluate performance using the standard F-measure. The labels for this task are shown in the top right of Figure \ref{fig:example}.

\textbf{4. Error Correction}: output the original, un-degraded version of each excerpt. In Figure \ref{fig:example}, a model is given the degraded excerpt (right) and expected to output the original excerpt (left). For this task, we define our own metric, helpfulness ($H$), based on two F-measures proposed by \cite{Bay2009}: frame-based F-measure with 40ms frames, and note-based onset-only F-measure. We use the \mbox{\texttt{mir\_eval}} \cite{mireval} implementation of note-based F-measure (with 50ms onset tolerance) to evaluate both the given excerpt and the system's output compared to the original excerpt. We take the average between the two F-measures for each excerpt, which we denote $F_g$ (for the given excerpt) and $F_c$ (for the system's corrected output). If $F_g = 1$ (the given excerpt was not degraded), $H = F_c$. If the given excerpt was degraded, however ($F_g < 1$), $H$ is calculated as in Equation (\ref{eq:metric}).
An intuition for this calculation is as follows: $H = 0.5$ represents an output which is exactly as accurate as the given excerpt (the error correction system has neither helped nor hurt), and $H$ scales linearly up to $1$ and down to $0$ from there. For example, $H = 0.75$ represents an output which is in some sense twice as accurate as the given excerpt (its error, $1 - F_c$, is half of the given excerpt's error, $1 - F_g$). Similarly, $H = 0.25$ represents an output which has twice as many errors as the input.
\begin{equation}
    H =
    \begin{cases}
        1 - \frac{1}{2}\frac{1 - F_c}{1 - F_g} & F_c \geq F_g \\
        \frac{1}{2}\frac{F_c}{F_g} & F_c < F_g
    \end{cases}
\label{eq:metric}
\end{equation}

\subsection{Baseline Models}
\subsubsection{Data Formats}
For input into our baseline models, we first quantize each excerpt onto 40 ms frames, rounding note onsets and offsets to the nearest frame. We use two different input formats for our baseline models, and provide data conversion and loading functions for each of them. 

The \textit{command format} is based on the one designed by \cite{oore2020time}. Each excerpt is converted into a sequence of one-hot vectors representing commands from a pre-defined vocabulary of 356 commands: $note\_on(p)$, $note\_off(p)$, and $shift(t)$ ($p \in [0, 127]$, $t \in [1, 100]$). The note on and off commands represent note onsets and offsets at the current frame, and the shift command skips $t$ frames. Longer shifts are represented by multiple shift commands.

The \textit{piano roll format} represents each excerpt as two binary piano rolls: one representing pitch presence in each frame, and another represent pitch onsets in each frame. These two piano rolls are concatenated together frame-wise to form the model's input.


\subsubsection{Model details}
The details for the models provided in this paper are brief. For code which fully defines the models and the code used to train and evaluate them, see the repo\footnote{\url{\mdtkurl}}. Our choice of models was relatively arbitrary; they are easy to implement with existing open source packages and easy to improve upon.

Our baseline for \textbf{Error Detection} uses the \textit{command format} as input. It consists of an embedding layer of size 128, followed by a basic Long Short-Term Memory (LSTM) \cite{hochreiter1997long}. A dropout of $0.1$ is applied to the final LSTM state's output, which is then passed to a fully-connected layer of size 2 with softmax activation, resulting in a single output for each input sequence.

Our \textbf{Error Classification} baseline uses the same design, but with output dimensionality 9 for the final layer (one for each degradation plus one for no degradation).

For \textbf{Error Location}, we use the \textit{piano roll format}. We first feed the input frames into a bi-directional LSTM (Bi-LSTM), and send the output of each Bi-LSTM state (with dropout $0.1$) into 3 linear layers, each with dropout $0.1$ and ELU activation. These are each fed into a final fully-connected layer of size 2 with softmax activation, resulting in one output per input frame.

For \textbf{Error Correction}, we use the \textit{piano roll format}, and base our network on a basic Encoder-Decoder structure \cite{cho2014learning}, where both the encoder and the decoder are Bi-LSTMs. The input is passed directly into the encoder Bi-LSTM, and the output at each frame is passed through a single fully connected layer with dropout $0.1$. This sequence is input into the decoder Bi-LSTM, each output of which is fed into 4 linear layers which output a vector of the same length as the input.

The models were trained using the Adam optimizer \cite{adam}, and a grid search was performed for weight decay, learning rate, LSTM hidden-unit size, and linear layer sizes (for full details, see the code). The model with the lowest validation loss on each task is used as the baseline.


\subsection{Analysis}
To gauge the difficulty of each task, we compare each of the baseline models to a simple rule-based approach. Like our baseline models, the rule-based systems output probability values $\in [0, 1]$. For Error Detection, the rule-based system returns an $\frac{8}{9}$ probability of each data point being degraded. For Error Classification, the rule-based system outputs a $\frac{1}{9}$ probability for each class. For Error Location, the rule-based system outputs a $0.06$ probability that each frame has been degraded (the proportion of frames that are degraded in the training set is $0.06$). Finally, for Error Correction, we calculate $p(1|0)=0.01$ and $p(1|1)=0.96$ from the training set\footnote{That is, for the degraded piano rolls from the training set, $1\%$ of cells with a $0$ and $96\%$ of cells with a $1$ map to a value of $1$ in the corresponding cell of the clean piano roll.} and have the system output these values for each cell in a given piano roll.

The results for each task on the ACME v1.0 test set are shown in Table \ref{tab:results}. From the losses, it is clear that the baseline models have learned something, since all of their losses are lower than the rule-based losses except for in Error Correction. However, from the metrics, it is also clear that there is much room for improvement on each of the proposed tasks (as we would hope).

\begin{table}
\centering
\begin{tabular}{llrr}
\toprule
Task & Model & Loss & Metric \\
\midrule
Error Detection      & Rule-based &  0.466 &      0.000 \\
                     & Baseline &  \textbf{0.344} &      0.000 \\
                     \midrule
Error Classification & Rule-based &  2.197 &  0.113 \\
                     & Baseline &  \textbf{2.130} &  \textbf{0.189} \\
                     \midrule
Error Location       & Rule-based &  0.404 &      0.000 \\
                     & Baseline &  \textbf{0.109} &  \textbf{0.525} \\
                     \midrule
Error Correction     & Rule-based &  \textbf{0.690} &   \textbf{0.590} \\
                     & Baseline &  0.693 &      0.000 \\
\bottomrule
\end{tabular}
    \caption{Loss and evaluation metric for the baseline and rule-based models for each task on the ACME v1.0 test set. Each task's metric is different, as explained in the text.}
    \label{tab:results}
\end{table}

For Error Detection, the baseline predicts 1 (degraded) for every data point, just like the rule-based system, likely because of the skew of the training data. As a simple attempt to overcome this tendency, we trained another model identical to the baseline which weights the loss of each data point inversely proportional to that label's frequency in the training set. This results in a model with greater overall loss (as expected), but which outputs some 0s, achieving a reverse F-measure of 0.155. Overcoming the skew of the dataset may prove to be a challenge for this task.

\begin{figure}
    \centering
    \includegraphics[width=0.45\textwidth,trim={0.3cm 3.15cm 0.5cm 1.05cm},clip]{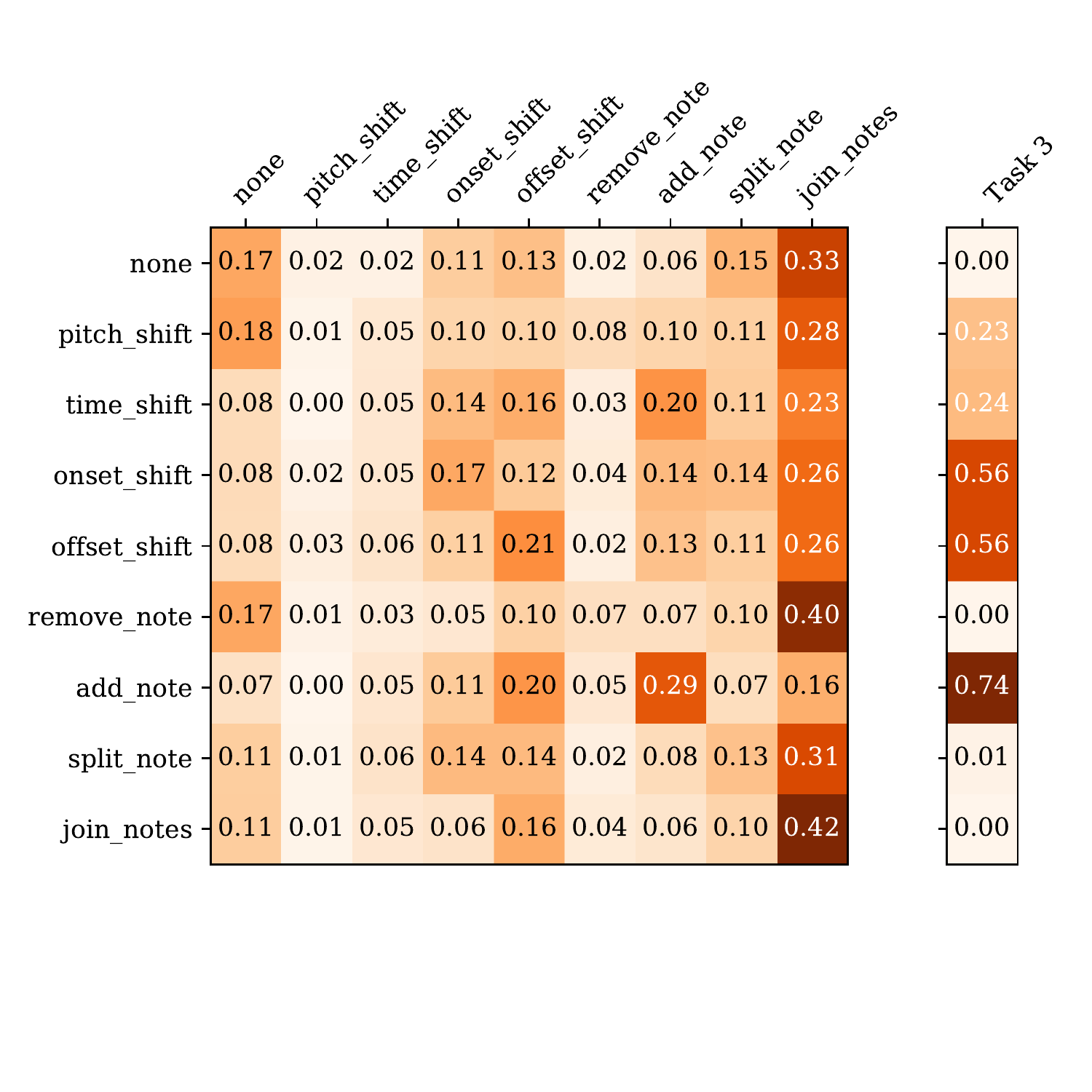}
    \caption{Left: Confusion matrix showing the distribution of the baseline Error Classification model's classifications, normalized by true label. Rows show the true label, and columns show the predicted label. Right: The baseline Error Location model's F-measure for each degradation type.}
    \label{fig:confmat}
\end{figure}

For Error Classification, the baseline achieves an accuracy of greater than that of the rule-based system. The baseline's confusion matrix is shown in Figure \ref{fig:confmat} (left), where rows represent the ground truth label and columns represent its output. This shows error tendencies, and (more importantly) gives an idea of the general difficulty of detecting each degradation. Here, it can be seen that the maximum point in each column is always on the diagonal, showing that the model does seem to have learned something sensible. It performs well on the add note degradation, classifying $32\%$ of those data points correctly. Pitch shift, time shift, and remove note seem to be the most difficult, while join notes is a common target for false positives. We are interested to see whether the above trends continue in future work on Error Classification, and intend to further investigate their causes.

The Error Location baseline outperforms the rule-based system in terms of both loss and F-measure by wide margins. It achieves this F-measure with a precision of 0.844 and a recall of 0.381, so although it rarely guesses that a frame has been degraded, it is usually correct when it does. Figure \ref{fig:confmat} (right) presents the baseline's F-measure split by degradation type, which shows the model performing best on add\_note, but also well for onset and offset shifts (precision is over $0.9$ for all three). It is slightly worse with pitch and time shifts (precision over $0.6$ for both), and performs poorly on the other degradations (the value for ``none'' will always be 0 since it has no positives). Given the relative success of this model compared to the other tasks' baselines, pre-training a model for this task before continuing to train it for another task might be an avenue for improved performance. Another strategy could be to use a model trained for this task as an attention mechanism for some of the other tasks.

Error Correction is clearly the most difficult task of the four, and the baseline model's performance reflects this. Although its loss is similar to that of the rule-based system, its helpfulness lags clearly behind. The rule-based model's strategy of (essentially) reproducing the input turns out to be a strong baseline. Our baseline, on the other hand, almost always outputs empty piano rolls, no matter the input. The difficulty of this task might require a more modular approach than the presented end-to-end baseline, perhaps combining the results of models from tasks 2 and 3 with a system designed to correct a specific degradation affecting a specific set of frames.

\section{Conclusion}
\label{sec:conc}
In this paper, we have introduced the MIDI Degradation Toolkit (MDTK), which contains tools to ``degrade'' (introduce errors to) MIDI excerpts. The toolkit is publicly available online\footnote{\url{\mdtkurl}} under an MIT License, and we encourage contributions and feedback from the community. Using MDTK, we have created the Altered and Corrupted MIDI Excerpts v1.0 (ACME v1.0) dataset\footnote{\url{\acmeurl}} and include in MDTK a tool to create custom ACME-style datasets with different settings or data. We have proposed a set of four new tasks of increasing difficulty involving such datasets: Error Detection, Classification, Location, and Correction, and designed evaluation metrics and scripts for each of them. We also designed and presented simple models to be used as a baseline for each, which show that the proposed tasks are non-trivial, and may require innovative solutions.

The toolkit is ready to be used for improving Automatic Music Transcription (AMT). To do so, a user can:
\begin{enumerate}[nolistsep]
    \item use \texttt{\mbox{measure_errors.py}} to analyse the types of errors made by an AMT system or acoustic model.
    \item instantiate a \texttt{\mbox{Degrader}} with the configuration produced by \texttt{\mbox{measure_errors.py}}---this can generate unlimited data matching the errors made by the system from step (1).
    \item train a discriminative model using data generated by the \texttt{\mbox{Degrader}}.
    \item apply that model to the output of the model from step (1) and evaluate the difference in performance.
\end{enumerate}

As performance on the proposed tasks modelling ACME v1.0 improves, we intend to introduce ACME v2.0 with additional features such as multi-track excerpts, a track-based degradation, longer excerpts, multiple degradations per excerpt, and various parameter settings for the degradations. We also intend to analyze the effect of adding noise on MLM performance.

\section{Acknowledgements}
Authors 1 and 2 contributed equally to this work.
This work is supported in part by JST ACCEL No. JPMJAC1602 and JSPS KAKENHI Nos. 16H01744 and 19H04137.

\bibliography{refs}

\end{document}